\documentclass[fleqn,10pt]{wlscirep}
 \pdfoutput=1
\usepackage[utf8]{inputenc}
\usepackage[T1]{fontenc}
\usepackage{graphicx}
\usepackage{multirow}
\usepackage{longtable}
\usepackage{makecell}
\usepackage{amsmath}
\usepackage{babel}
\usepackage{comment}
\usepackage[font=small,labelfont=bf]{caption}
\title{Predicting potential drug targets and repurposable drugs for COVID-19 via a deep generative model for graphs}

\author[1*,+]{Sumanta Ray}
\author[2+]{Snehalika Lall}
\author[3]{Anirban Mukhopadhyay}
\author[2]{Sanghamitra Bandyopadhyay}
\author[1,4*]{Alexander Schönhuth}
\affil[1]{Centrum Wiskunde \& Informatica, Science Park 123, 1098 XG Amsterdam, The Netherlands}
\affil[2]{Machine Intelligence Unit, Indian Statistical Institute, Kolkata, India.}
\affil[3]{Department of Computer Science and Engineering, University of Kalynai, Kalyani, India}
\affil[4]{Genome Data Science, Bielefeld University, Bielefeld, Germany}

\affil[*]{Sumanta.Ray@cwi.nl}

\affil[+]{these authors contributed equally to this work}

\begin{abstract}

Coronavirus Disease 2019 (COVID-19) has been creating a worldwide pandemic situation. Repurposing drugs, already shown to be free of harmful side effects, for the treatment of COVID-19 patients is an important option in launching novel therapeutic strategies. Therefore, reliable molecule interaction data are a crucial basis, where drug-/protein-protein interaction networks establish invaluable, year-long carefully curated data resources. However, these resources have not yet been systematically exploited using high-performance artificial intelligence approaches. Here, we combine three networks, two of which are year-long curated, and one of which, on SARS-CoV-2-human host-virus protein interactions, was published only most recently (30th of April 2020), raising a novel network that puts drugs, human and virus proteins into mutual context. We apply Variational Graph AutoEncoders (VGAEs), representing most advanced deep learning based methodology for the analysis of data that are subject to network constraints. Reliable simulations confirm that we operate at utmost accuracy in terms of predicting missing links. We then predict hitherto unknown links between drugs and human proteins against which virus proteins preferably bind. The corresponding therapeutic agents present splendid starting points for exploring novel host-directed therapy (HDT) options.

\end{abstract}

\begin{document}
\flushbottom
\maketitle
%
%
\thispagestyle{empty}


\section*{Introduction}
The pandemic of COVID-19 (Coronavirus Disease-2019) has affected more than 6 million people. So far, it has caused about 0.4 million deaths in over 200 countries worldwide (https://coronavirus.jhu.edu/map.html), with numbers still increasing rapidly. COVID-19 is an acute respiratory disease caused by a highly virulent and contagious novel coronavirus strain, SARS-CoV-2, which is an enveloped, single-stranded RNA virus \cite{wu2020new}. Sensing the urgency, researchers have been relentlessly searching for possible therapeutic strategies in the last few weeks, so as to control the rapid spread.

In their quest, drug repurposing establishes one of the most relevant options, where drugs that have been approved (at least preclinically) for fighting other diseases, are screened for their possible alternative use against the disease of interest, which is COVID-19 here. Because they were shown to lack severe side effects before, risks in the immediate application of repurposed drugs are limited. In comparison with de novo drug design, repurposing drugs offers various advantages. Most importantly, the reduced time frame in development suits the urgency of the situation in general. Furthermore, most recent, and most advanced artificial intelligence (AI) approaches have boosted drug repurposing in terms of throughput and accuracy enormously. Finally, it is important to understand that the 3D structures of the majority of viral proteins have remained largely unknown, which raises the puts up the obstacles for direct approaches to work even higher.

The foundation of AI based drug repurposing are molecule interaction data, optimally reflecting how drugs, viral and host proteins get into contact with each other. During the life cycle of a virus, the viral proteins interact with various human proteins in the infected cells. Through these interactions, the virus hijacks the host cell machinery for replication, thereby affecting the normal function of the proteins it interacts with. To develop suitable therapeutic strategies and design antiviral drugs, a comprehensive understanding of the interactions between viral and human proteins is essential \cite{forst2010host}. 

When watching out for drugs that can be repurposed to fight the virus, one has to realize that targeting single virus proteins easily leads to the viruses escaping the (rather simpleminded) attack by raising resistance-inducing mutations. Therefore, host-directed therapies (HDT), which target human proteins that represent important carriers for the virus to enter and manipulate the human cells, offer an important supplementary strategy \cite{kaufmann2018host}. Unlike strategies that directly target the proteins of the virus, HDT are thought to be less prone to developing resistance, because human proteins are less affected by mutations, and therefore represent more sustainable drug targets. For establishing HDT, one has to identify proteins that are crucial for maintenance and perseverance of the disease causing virus in the human cells. Once such proteins are targeted, the replication machinery of the virus falls apart.

For all these reasons, repurposing drugs for HDT against COVID-19 has great potential. Moreover, it provides hope for rapid practical implementation because of the lack of side effects.

Because the basis for drug repurposing screens are molecule interaction data, biological interaction networks offer invaluable resources, because they have been carefully curated and steadily refined for many years. This immediately points out that network based repurposing screens offer unprecedented opportunities in revealing targets for HDT's, which explains the recent popularity of such approaches in general\cite{alaimo2019network}. Evidence for the opportunities is further provided by successful approaches based on viral-host networks in particular\cite{de2012new,emig2013drug}, having led to therapy options for treating Dengue \cite{doolittle2011mapping}, HIV \cite{bandyopadhyay2015review}, Hepatitis C \cite{mukhopadhyay2014network} and Ebola \cite{cao2017prediction}. Beyond fighting viruses, treatments for various other diseases have been developed\cite{cheng2019genome,zeng2019deepdr}.

Since the outbreak of COVID-19, a handful of research groups have been trying to exploit network resources, developing network algorithms for the discovery of drugs that can be repurposed to act against SARS-CoV-2. The first attempt was made by Zhou et al.\cite{zhou2020network} through an integrative network analysis, followed by Li et al.\cite{li2020network}, who combined network data with a comparative analysis on the gene sequences of different viruses to obtain drugs that can potentially be repurposed to act against SARS-CoV-2.

Shortly thereafter, Gordon et al.\cite{gordon2020sars} conducted pioneering work by generating a map that juxtaposes SARS-CoV-2 proteins with human proteins that were found to interact in affinity-purification mass spectrometry (AP-MS) screens. Furthermore, in independent work, Dick et al.\cite{SP2dick} identified high confidence interactions between human proteins and SARS-CoV-2 proteins using sequence-based PPI predictors (a.k.a.~PIPE4 \& SPRINT).

Both of the studies\cite{gordon2020sars,SP2dick} on SARS-CoV-2 provide data that had so far been urgently missing in the fight against COVID-19. Only now, we are able to link existing (long term curated and highly reliable) drug-protein and human protein-protein interaction data with proteins of SARS-CoV-2. In other words, only now we can draw links between the molecular agents of the virus and existing drugs on a scale that is sufficiently large to allow for systematic high throughput repurposing screens. Still, of course, it remains to design, develop and implement the corresponding strategies in order to exploit the now decisively augmented resources.

As above-mentioned, in order to exploit resources to a maximum, one optimally makes use of sufficiently advanced AI techniques. In this work, to the best of our knowledge for the first time, we combine all arguments raised above.\\

{\bf (1)} We link existing high-quality, long-term curated and refined, large scale drug/protein - protein interaction data with

{\bf (2)} molecular interaction data on SARS-CoV-2 itself, raised only a handful of weeks ago,

{\bf (3)} exploit the resulting overarching network using most advanced, AI boosted techniques

{\bf (4)} for repurposing drugs in the fight against SARS-CoV-2

{\bf (5)} in the frame of HDT based strategies.\\

As for (3)-(5), we will highlight interactions between SARS-Cov-2-host protein and human proteins important for the virus to persist using most advanced deep learning techniques that cater to exploiting network data. We are convinced that many of the fairly broad spectrum of drugs we raise will be amenable to developing successful HDT's against COVID-19. 

\begin{figure*}[!h]
\centering
   \includegraphics[width=0.99\linewidth]{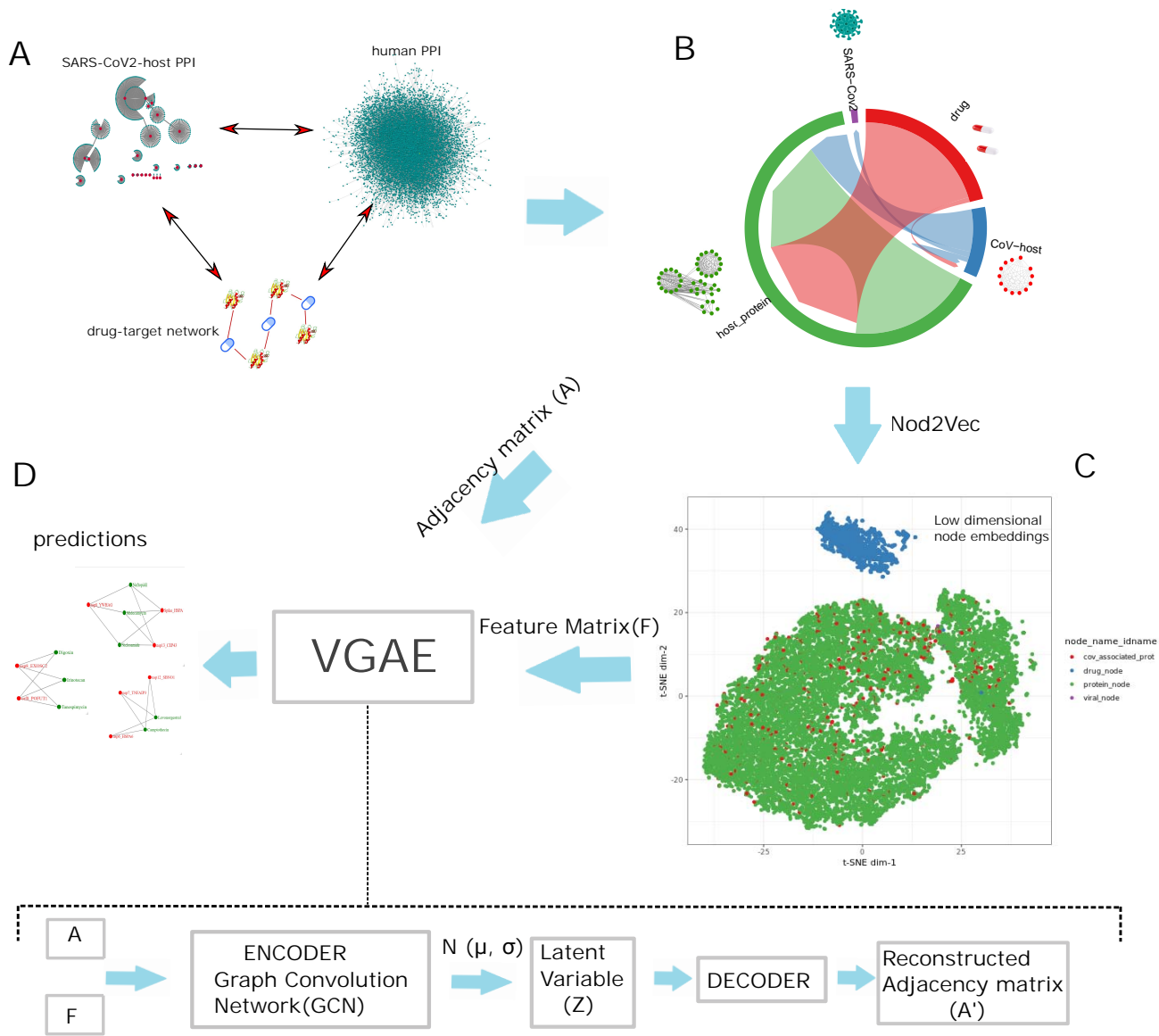}
\caption{Overall workflow of the proposed method: The three networks SARS-CoV-2-host PPI, human PPI, and drug-target network (Panel-A) are mapped by their common interactors to form an integrated representation (Panel-B). The neighborhood sampling strategy Node2Vec converts  the network into fixed-size low dimensional representations that perverse the properties of the nodes belonging to the three major components of the integrated network (Panel-C). The resulting feature matrix (F) from the node embeddings and adjacency matrix (A) from the integrated network are used to train a VGAE model, which is then used for prediction (panel-D). }
\label{fig:framework}
\end{figure*}

\section*{Results}

In the following, we will first describe the workflow of our analysis pipeline and the basic ideas that support it.

We proceed by carrying out a simulation study that proves that our pipeline accurately predicts missing links in the encompassing drug - human protein - SARS-CoV-2-protein network that we raise and analyze. Namely we demonstrate that our (high-performance, AI supported) prediction pipeline accurately re-establishes links that had been explicitly removed before. This provides sound evidence that the interactions that we predict in the full network most likely reflect true interactions between molecular interfaces.

Subsequently, we continue with the core experiments. We predict links to be missing in the full (without artificially having removed links), encompassing drug - human protein - SARS-CoV-2-protein network, raised by combining links from year-long curated resources on the one hand and most recently published COVID-19 resources on the other hand. As per our simulation study, a large fraction, if not the vast majority of the predictions establish true, hence actionable interactions between drugs on the one hand and SARS-CoV-2 associated human proteins (hence of use in HDT) on the other hand.

For the purposes of high-confidence validation, we carry out a literature study on the overall 92 drugs we put forward. For this, we inspect the postulated mechanism-of-action of the drugs in the frame of several diseases, including SARS-CoV and MERS-CoV driven diseases in particular. 

\subsection*{Workflow}

See Figure~\ref{fig:framework} for the workflow of our analysis pipeline and the basic ideas that support it. We will describe all important steps in the paragraphs of this subsection.

\paragraph{\em Raising a Comprehensive Interaction Network.}
See A \& B in Figure~\ref{fig:framework}. We have combined three interaction networks, two of which represent year-long curated and much refined publicly available resources, namely drug-gene interaction and the human interactome, together compiled from eight different, reliable and publicly accessible sources, and one of which the SARS-CoV-2--human protein-protein-interaction (PPI) network was published only a few weeks ago. The integrated network has four types of nodes:

{\bf 1)} SARS-CoV-2 proteins,

{\bf 2)} SARS-CoV-2-associated host proteins (CoV-host),

{\bf 3)} human proteins other than 2)

{\bf 4)} and drugs.\\
This means that we put drugs, human proteins and SARS-CoV-2 proteins into mutual context via the links provided by this network. However, because the encompassing network is built from three subnetworks, links between nodes from different individual subnetworks are presumably missing. It remains to predict them using an AI approach, preferably of utmost performance. This AI approach needs to identify links across the individual subnetworks. Because new cross-subnetwork links may imply links within the established networks as a consequence, the AI approach should also predict new links within the individual parts of our network, if this is called for.   


\paragraph{\em AI Model First Stage: Node2Vec }
See C in Figure~\ref{fig:framework}. To bring the link prediction machinery into effect, we raise a model that operates in two stages. First, we employ a network embedding strategy (here: Node2Vec\cite{grover2016node2vec}), which extracts node features from the integrated network. In a bit more formal detail, Node2Vec converts the adjacency matrix that represents the network into a fixed-size, low-dimensional latent space the elements of which are the feature vectors of the nodes. Thereby, Node2Vec aims at preserving the properties of the nodes relative to their surroundings in the network. For efficiency reasons, Node2Vec makes use of a sampling strategy. The result of this step is a feature matrix ($F$) where rows refer to nodes and columns refer to the inferred network features.

\paragraph{\em AI Model Second Stage: Variational Graph Autoencoders (VGAE).}
See B, C \& D in Figure~\ref{fig:framework}. In the next step, we employ variational graph autoencoders (VGAE), as a most recent, graph neural network based technique shown to be of utmost accuracy, to predict links in networks that although missing, are highly likely to exist\cite{kipf2016variational}. VGAE's require the original graph, provided by its adjacency matrix $A$ and, optionally, a feature matrix that annotates the nodes of the network with helpful additional information. Often, $F$ does not necessarily refer to the topology of the network itself. Here, however, we do make use of the feature matrix $F$ that was inferred from $A$ itself in the first step. We found that, despite just being an alternative representation of $A$, using $F$ aided in raising prediction accuracy substantially. This may not be surprising, however, because $F$ consists of knowledge obtained using Node2Vec, which, as being complementary to VGAE's from a methodical point of view, cannot necessarily revealed by VGAE's itself. 

\paragraph{\em Predicting Missing Links.}
See D in Figure~\ref{fig:framework}. After training the VGAE, we finally predict links in the encompassing drug-human-virus interaction network that had remained to be missing. For this, we make use of the decoding part of the VGAE, which re-raises the network based on the latent representation the network provided by the encoder. Re-raising the network results in edges between nodes that although not having been explicit before, are imperative to exist relative to the encoded version of the network. Thereby, one predicts links between drugs and SARS-CoV-2-associated human proteins in particular. Although these links had not been explicit elements of the drug-human interaction subnetwork before, their existence is implied by the topological constraints the comprehensive network imposes. Thus, our model predicts both drugs and proteins: repurposing these drugs leads to them targeting the matching proteins. See Figure~\ref{fig:framework} for the total workflow we just described.


\paragraph{\em Addressing Computation Time.}
To reduce the computation time, we used a fast version of VGAE's as proposed by Salha et al., \cite{salha2020fastgae}. This fast version relies on a strategy by which to sample nodes. Using several non-overlapping test sets for different numbers of sampling nodes, we evaluated the model based on sampling 5000 nodes as most suitable for our prediction task.


\subsection*{Validation of the Learned Model}
Let $G=(V,E)$ be the entire drug-human-virus interaction network in the following, where nodes $V$ represent drugs or proteins and edges $E$ represent interactions between them. We run Node2Vec\cite{grover2016node2vec} on $G$ to obtain a feature matrix $F$ where rows can be identfied with elements from $V$ and columns represent features extracted from $G$.

After having computed $F$, we encode the network $G$ into the embedding space using variational graph autoencoder (VGAE) techniques. In detail, we use the FastGAE model\cite{salha2020fastgae} as a version that decisively speeds up the learning process in the decoding phase as a major argument. For encoding, FastGAE utilizes a (popular) Graph Convolutional Network (GCN) encoder. This leads to encoding all nodes into the latent variable based embedding space. Therefore, FastGAE makes use of the original graph adjacency matrix $A$, which codes the original topology of the graph, and $F$. Using FastGAE is explained by the fact that in the decoding phase, a new version of the adjacency matrix has to be established, where $A$ is huge ($N\times N$, where $N$ is the number of all drugs, human proteins and SARS-CoV-2 proteins together), which slows down less rapid implementations of VGAE's decisively. To sort out this issue, FastGAE randomly samples subgraphs $G_S$, referring to smaller sets of nodes $S\subset V$ of size $N_S$, and reconstructs the corresponding submatrices $A_S$ in several iterations, each of which refers to a different $S$. Finally, the submatrices of all samples are combined into an overarching matrix $\tilde{A}$, as an approximation of the matrix that gets reconstructed as a whole in the decoding phase of slow approaches. Note that the approximation was shown to be highly accurate\cite{salha2020fastgae}.


This reduces the training time compared to the general graph autoencoder model. We tested the model performance for a different number of sampled nodes, keeping track of the area under the ROC curve (AUC), average precision (AP) score, and model training time in the frame of a train-validation-test split at proportions 8:1:1.  Table \ref{perform_model} shows the performance of the model for sampled sugraph sizes $N_S$= 7000, 5000, 3000, 2500 and 1000. For 5000 sampled nodes, the model's performance is sufficiently good enough concerning its training time and validation-AUC and -AP score. The average test ROC-AUC and AP score of the model for $N_s$=5000 are $88.53 \pm 0.03$ and $84.44 \pm 0.04$.

To know the efficacy of the model in discovering the existing edges between only CoV-host and drug nodes, we train the model (with $N_s$=5000) on an incomplete version of the graph where the links between CoV-host and drugs have been removed. We further compute the feature matrix $F$ based on the incomplete graph, and use it. The test set consists of all the previously removed edges. The model performance is no doubt better for discovering those edges between CoV-host and drug nodes (ROC-AUC: $93.56 \pm 0.01$ AP: $90.88 \pm 0.02$ for 100 runs). 
 
The FastGAE model is learned with the feature matrix ($F$) and adjacency matrix ($A$). The node feature matrix ($F$) is obtained from $A$ using the Node2Vec neighborhood sampling strategy. The model performance is evaluated with and without using $F$ as feature matrix. Figure \ref{fig:withwithoutF} shows the average performance of the model on validation sets with and without $F$ as input for the different number of sampling nodes. We calculate average AUC, and AP scores for 50 complete runs of the model. From figure \ref{fig:withwithoutF}, it is evident that including $F$ as feature matrix enhances the model's performance markedly.


\begin{minipage}{\textwidth}
   \begin{minipage}[b]{\textwidth}
   \captionof{table}{Performance of the FastGAE model for different sampling nodes ($N_s$): mean validation-AUC and -AP scores is computed over last 10 epoch in each run with a train-validation-test split of 8:1:1. Performance is reported over 100 runs for $N_s$=7000, 5000, 3000, 2500 and 1000}
    \centering
  \begin{tabular}{|l|l|l|l|}
\hline
\multirow{2}{*}{$N_s$} & \multicolumn{3}{|c|}{Average Performance on Validation Set}  \\
\cline{2-4}
 & AUC ($\%$)& AP ($\%$) & Training Time (in sec) \\
\hline
7000 & 89.21 $\pm$ 0.02 & 85.32 $\pm$ 0.02 & 1587\\
\hline
5000 & 89.17 $\pm$ 0.03 & 85.30 $\pm$ 0.04 & 1259\\
\hline
3000 & 88.91 $\pm$ 0.10 & 85.02 $\pm$ 0.04 & 1026\\
\hline
2500 & 88.27 $\pm$ 0.15 & 84.88 $\pm$ 0.13 & 998\\
\hline
1000 & 86.69 $\pm$ 0.17 & 83.58 $\pm$ 0.19 & 816 \\
\hline

\end{tabular}
\label{perform_model}
    \end{minipage}
   
     \begin{minipage}[b]{\textwidth}
    \captionof{figure}{Performance of the model in validation set with and without using feature matrix (F)}
    \centering
    \includegraphics[scale=0.4]{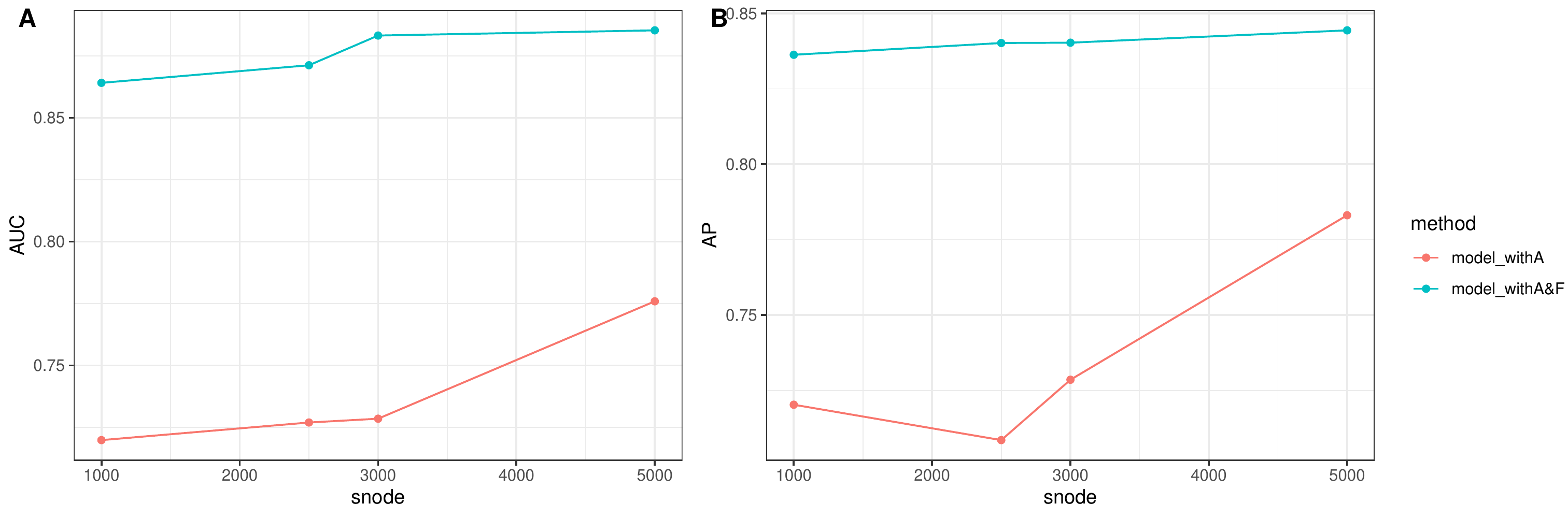}
       \label{fig:withwithoutF}
  \end{minipage}
  
  \end{minipage}

\begin{figure*}[t]
    \centering
    \includegraphics[width=\textwidth]{tsne_all_panel_new.pdf}
    \caption{Results of applying Node2Vec on the whole network. Panel-A shows 2-dimensional t-SNE plot of all nodes in the network. Panel-B shows the same for only drug and CoV-host nodes. Panel-C represents heatmap of 21 `most similar' (structurally equivalent or have similar `role') drugs of CoV-hosts. The drugs are colored based on their clinical phase (red--launched, preclinical--blue, phase2/phase3--green and phase-1/ phase-2--black ). Panel-D represents visualization of 17 Louvain clusters identified in the low dimensional embedding space.  Panel-E shows a network consisting of 6 drugs and their most probable neighbours within the network. These drugs share same Louvain cluster with some CoV-host proteins. }
    \label{fig:node2vec}
\end{figure*}

\subsection*{Graph Embedding of the Compiled Network}
We use the Node2Vec framework to learn low dimensional embeddings of each node in the compiled network. It uses the Skipgram algorithm of the word2vec model to learn the embeddings, which eventually groups nodes with a similar `role' or having a similar `connection pattern' within the graph. Similar `role' ensures that nodes within the sets/groups are structurally similar/equivalent than the other nodes outside the groups. Two nodes are said to be structurally equivalent if they have identical connection patterns to the rest of the network \cite{rossi2019community}. 
To explore this, we have analyzed the embedding results in two steps. First, we explore structurally equivalent nodes to identify 'roles' and similar connection patterns to the rest of the networks, and later use Lovain clustering to examine the same within the groups/clusters. The $'most\_similar'$ function of the Node2Vec inspects the structurally equivalent nodes within the network. We find out all the CoV-host nodes which are most similar to the drug nodes. While it is expected to observe nodes of the same types within the neighborhood of a particular node, in some cases, we found some drugs are neighbors of CoV-host proteins with high probability ($pobs >$ 0.65). Figure \ref{fig:node2vec} panel-C shows a heatmap of 21 such drugs and their most similar CoV-host proteins. The drugs are hierarchically clustered based on the similarity scores and demonstrated in the row-side dendrogram of the heatmap. A couple of these drugs are frequently used to treat severe allergies, asthma, respiratory diseases, and gastrointestinal infections, which are the main symptoms of COVID patients. Mainly, Dexamethasone, Cetirizine, isoprenaline, Betulinic acid and Clebopride, which are the probable neighbor of CoV-host EXOSC3 ( with $pobs$: 0.71, 0.69, 0.67, 0.68 and 0.68, respectively, figure \ref{fig:node2vec}, panel-C) have a well-known effect on these diseases. Betulinic acid has antiretroviral, antimalarial, and anti-inflammatory properties and acts as the inhibitors of SARS-CoV-2 3CL protease \cite{wen2007specific}. 
Some other drugs such as `Clenbuterol' and `Fenbendazole', the probable neighbor of ppp1cb and EEF1A respectively, are used as bronchodilators in asthma.

To explore the closely connected groups, we have constructed a neighborhood graph using the K-th nearest neighbor algorithm from the node embeddings and apply Louvain clustering (Figure \ref{fig:node2vec}-panel-C). Although there is a clear separation between host proteins (including CoV-host) cluster and drug cluster, some of the Louvain clusters contain both types of nodes. For example, Louvain cluster-16 and -17 contain four and two drugs along with the other CoV-host proteins, respectively. Figure \ref{fig:node2vec} panel-D represents a network consisting of these six drugs and their most similar CoV-host nodes. 

\begin{figure}
    \centering
    \includegraphics[width=\textwidth] {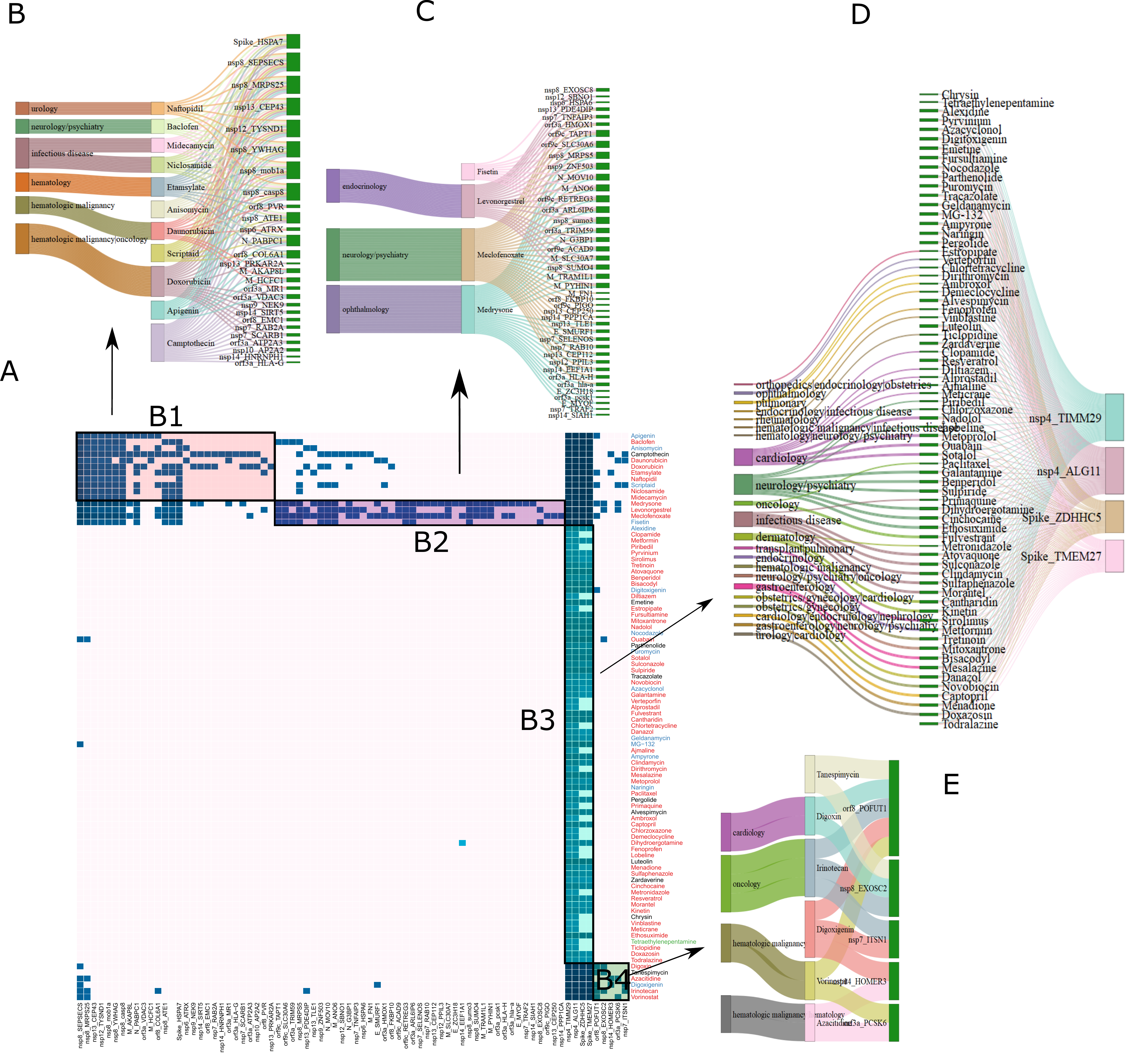}
    \caption{Drug--CoV-host predicted interaction: panel-A shows heatmap of probability scores between 92 drugs and 78 CoV-host proteins. The four predicted bipartite modules are annotated as B1, B2, B3 and B4 within the heatmap. The drugs are colored based on their clinical phase (red--launched, preclinical--blue, phase2/phase3--green and phase-1/ phase-2--black ). Panel-B, C, D and E represents networks corresponding to B1, B2, B3 and B4 modules.The drugs are annotated using the disease area found in CMAP database \cite{subramanian2017next}}
    \label{fig:shankey}
\end{figure}

\begin{figure}
    \centering
    \includegraphics[scale=0.09]{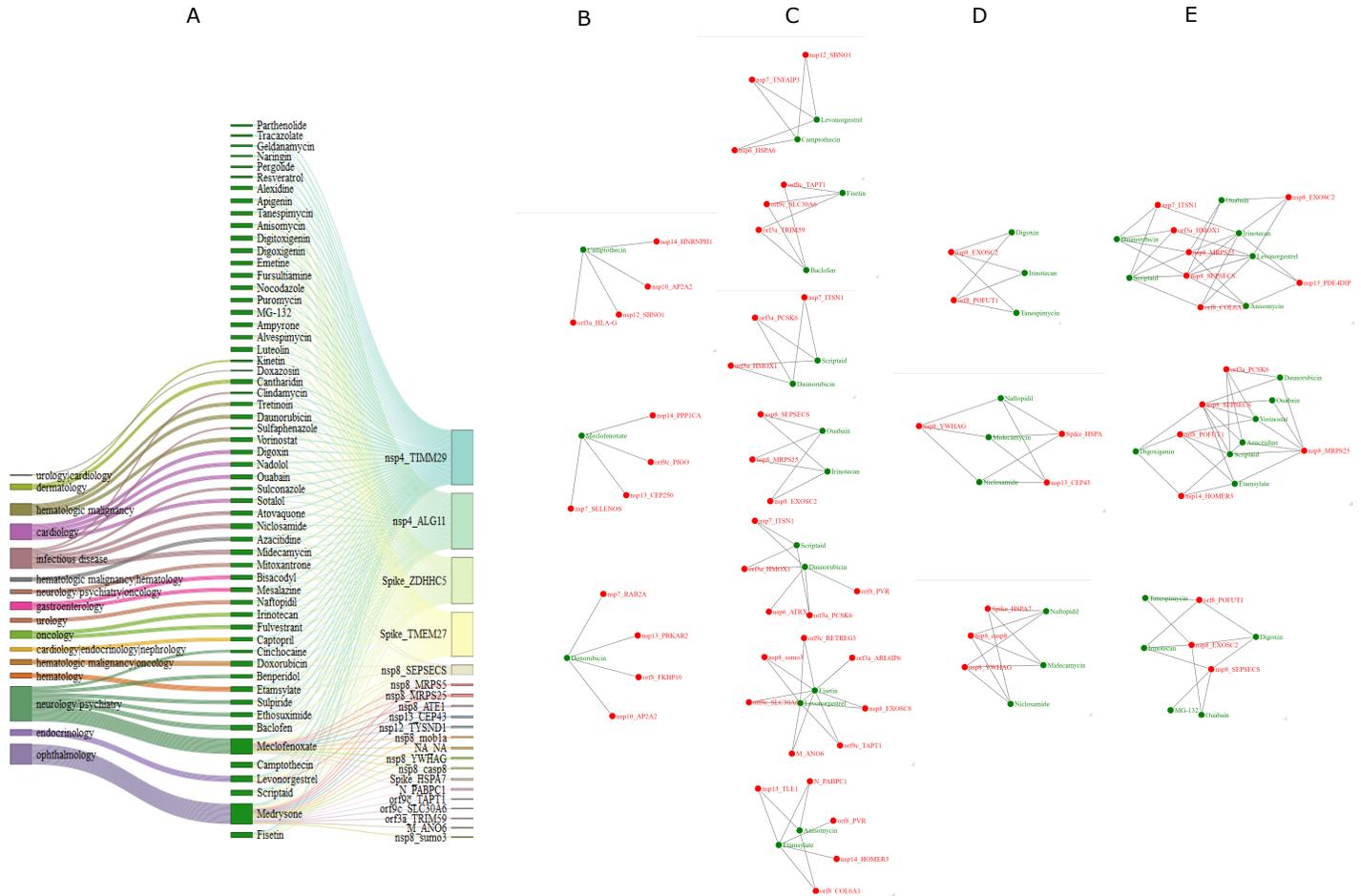}
    \caption{Predicted interactions for probability threshold: 0.9. panel-A shows the interaction graph between drugs and CoV-host. Drugs are annotated with their usage. Panel-B, C, D and E represents quasi-bicliques  for one, two, three and more than three drugs molecules respectively. }
    \label{fig:prediction0.9}
\end{figure}


\subsection*{Drug--CoV-host Interaction Prediction}
For drug--Cov-host interaction prediction, we exploit Variational Graph Autoencoder (VGAE), an unsupervised graph neural network model, first introduced in \cite{kipf2016variational} to leverage the concept of variational autoencoder in graph-structured data. To make learning faster, we utilized the fastGAE model to take advantage of the fast decoding phase. We have used two data matrices in the fastGAE model for learning: one is the adjacency matrix, which represents the interaction information over all the nodes, and the other one is the feature matrix representing the low-dimensional embeddings of all the nodes in the network. We create a test set of `non-edges' by removing all existing links between drugs and CoV-host proteins from all possible combinations (332 CoV-host $\times$ 1302 drugs) of edges. The model is trained on the whole network with the adjacency matrix $A$ and feature matrix $F$. The trained model is then applied to the test `non-edges' to know the most probable links. We identified a total of 692 most probable links with 92 drugs and 78 CoV-host proteins with a probability threshold of 0.8. The predicted CoV-host proteins are involved in different crucial pathways of viral infection (table \ref{tab:GOtable}). The p-values for pathway and GO enrichment are calculated by using the hypergeometric test with 0.05 FDR corrections. Figure~\ref{fig:shankey}, Panel-A shows the heatmap of probability scores between predicted drugs and CoV-host proteins. To get more details of the predicted bipartite graph, we use a weighted bipartite clustering algorithm proposed by J. Beckett \cite{beckett2016improved}. This results in 4 bipartite modules (Panel-A figure \ref{fig:shankey}): B1 (11 drugs, 28 CoV-host), B2 (4 drugs, 41 CoV-host), B3 (71 rugs and 4 CoV-host), and B4 ( 6 drugs and 5 CoV-host). The other panels of the figure show the network diagram of four bipartite modules. 
B1 contains 11 drugs, including some antibiotics (Anisomycin, Midecamycin), and anti-cancer drugs (Doxorubicin, Camptothecin). B3 also has some antibiotics such as Puromycin, Demeclocycline, Dirithromycin, Geldanamycin, and Chlortetracycline, among them, the first three are widely used for bronchitis, pneumonia, and respiratory tract infections \cite{wishart2006drugbank}. Some other drugs such as Lobeline and Ambroxol included in the B3 module have a variety of therapeutic uses, including respiratory disorders and bronchitis.
The high confidence predicted interactions (with threshold 0.9) is shown in Figure~\ref{fig:prediction0.9} panel-A. To highlight some repurposable drug combination and their predicted CoV-host target, we perform a weighted clustering (clusterONE) \cite{nepusz2012detecting} on this network and found some quasy-bicluques (shown in Panel-B-E) 

We matched our predicted drugs with the drug list recently published by Zhou et al. \cite{zhou2020network} and found six common drugs: Mesalazine, Vinblastine, Menadione, Medrysone, Fulvestrant, and Apigenin. 
Among them, Apigenin has a known effect in the antiviral activity together with quercetin, rutin, and other flavonoids \cite{salehi2019therapeutic}.  Mesalazine is also proven to be extremely effective in the treatment of other viral diseases like influenza A/H5N1 virus. \cite{zheng2008delayed}.    

\subsection*{Repurposable drugs for SARS-CoV-2}

\subsubsection*{Baclofen and  Fisetin}
Baclofen, a benzodiazepine receptor (GABAA-receptor) agonist, has a potential role in antiviral associated treatment \cite{leggio2012baclofen}. Anti-inflammatory antecedents fisetin is also tested for antiviral activity, such as for inhibition of Dengue (DENV) virus infection \cite{jasso2019antiviral}. It down-regulates the production of proinflammatory cytokines induced by a DENV infection. Both of the drugs are listed in the high confidence interaction set with the three CoV-hosts: TAPT1 (interacted with SARS-CoV-2 protein: orf9c), SLC30A6 (interacted with SARS-CoV-2 protein: orf9c), and TRIM59 (interacted with SARS-CoV-2 protein: orf3a) (Figure~\ref{fig:prediction0.9}-panel-C). 

\subsubsection*{Topoisomerase Inhibitors}
 Topoisomerase Inhibitors play an active role as antiviral agents by inhibiting the viral DNA replication \cite{maschera1993evaluation,gonzalez2012potent}. Some Topoisomerase Inhibitors such as Camptothecin, Daunorubicin, Doxorubicin, Irinotecan and Mitoxantrone are predicted to interact with several CoV-host proteins. It has been demonstrated that the anticancer drug camptothecin (CPT) and its derivative Irinotecan have a potential role in antiviral activity \cite{horwitz1972antiviral,bennett2016analog}. It inhibits host cell enzyme topoisomerase-I which is required for the initiation as well as completion of viral functions in host cell \cite{pantazis1999water}. Daunorubicin (DNR) has also been demonstrated as an inhibitor of HIV-1 virus replication in human host cells \cite{filion1993inhibition}. The conventional anticancer antibiotic Doxorubicin was identified as a selective inhibitor of {\em in vitro} Dengue and Yellow Fever virus replication \cite{kaptein2010derivate}. It is also reported that doxorubicin coupling with monoclonal antibody can create an immunoconjugate that can eliminate HIV-1 infection in mice cell \cite{johansson2006elimination}. Mitoxantrone shows antiviral activity against the human herpes simplex virus (HSV1) by reducing the transcription of viral genes in many human cells that are essential for DNA synthesis \cite{huang2019antiviral}.

 \subsubsection*{Histone Deacetylases Inhibitors (HDACi)}
 
 Histone Deacetylases Inhibitors (HDACi) are generally used as latency-reversing agents for purging HIV-1 from the latent reservoir like CD4 memory cell \cite{matalon2011histone}. Our predicted drug list (Table~\ref{drug_tab}) contains two HDACi: Scriptaid and Vorinostat. Vorinostrate can be used to achieve latency reversal in the HIV-1 virus safely and repeatedly \cite{archin2017interval}. 
 Asymptomatic patients infected with SARS-CoV-2 are of significant concern as they are more vulnerable to infect large number of people than symptomatic patients. Moreover, in most cases (99 percentile), patients develop symptoms after an average of 5-14 days, which is longer than the incubation period of SARS, MERS, or other viruses \cite{lauer2020incubation}. To this end, HDACi may serve as good candidates for recognizing and clearing the cells in which SARS-CoV-2 latency has been reversed.     

\subsubsection*{HSP inhibitor}
Heat shock protein 90 (HSP) is described as a crucial host factor in the life cycle of several viruses that includes an entry in the cell, nuclear import, transcription, and replication \cite{ju2011synthesis,shim2011heat}. HSP90 is also shown to be an essential factor for SARS-CoV-2 envelop (E) protein \cite{dediego2011severe}. In \cite{wang2017hsp90}, HSP90 is described as a promising target for antiviral drugs. The list of predicted drugs contains three HSP inhibitors: Tanespimycin, Geldanamycin, and its derivative Alvespimycin. The first two have a substantial effect in inhibiting the replication of Herpes Simplex Virus and Human enterovirus 71 (EV71), respectively. Recently in \cite{sultan2020drug}, Geldanamycin and its derivatives are proposed to be an effective drug in the treatment of COVID-19. 


\subsubsection*{Antimalarial agent|DNA inhibitor, DNA methyltransferase inhibitor, DNA synthesis inhibitor}

Inhibiting DNA synthesis during viral replication is one of the critical steps in disrupting the viral infection. The list of predicted drugs contains six such small molecules/drugs, viz., Niclosamide, Azacitidine, Anisomycin, Novobiocin, Primaquine, Menadione, and Metronidazole. DNA synthesis inhibitor Niclosamide has a great potential to treat a variety of viral infections, including SARS-CoV, MERS-CoV, and HCV virus \cite{xu2020broad} and has recently been described as a potential candidate to fight the SARS-CoV-2 virus \cite{xu2020broad}. Novobiocin, an aminocoumarin antibiotic, is also used in the treatment of Zika virus (ZIKV) infections due to its protease inhibitory activity. In 2005, Chloroquine (CQ) had been demonstrated as an effective drug against the spread of severe acute respiratory syndrome (SARS) coronavirus (SARS-CoV). Recently Hydroxychloroquine (HCQ) sulfate, a derivative of CQ, has been evaluated to efficiently inhibit SARS-CoV-2 infection {\em in vitro} \cite{liu2020hydroxychloroquine}. Therefore, another anti-malarial aminoquinolin drug Primaquine may also contribute to the attenuation of the inflammatory response of COVID-19 patients. Primaquine is also established to be effective in the treatment of Pneumocystis pneumonia (PCP) \cite{vohringer1993pharmacokinetic}.  

\subsubsection*{Cardiac Glycosides ATPase Inhibitor}
Cardiac glycosides have been shown to play a crucial role in antiviral drugs. These drugs target cell host proteins, which help reduce the resistance to antiviral treatments. The antiviral effects of cardiac glycosides have been described by inhibiting the pump function of Na, K-ATPase. This makes them essential drugs against human viral infections. The predicted list of drugs contains three cardiac glycosides ATPase inhibitors: Digoxin, Digitoxigenin, and Ouabain. These drugs have been reported to be effective against different viruses such as herpes simplex, influenza, chikungunya, coronavirus, and respiratory syncytial virus \cite{amarelle2018antiviral}.

\subsubsection*{MG132, Resveratrol and Captopril}
MG132, proteasomal inhibitor is established to be a strong inhibitor of SARS-CoV replication in early steps of the viral life cycle \cite{schneider2012severe}. MG132 inhibits the cysteine protease m-calpain, which results in a pronounced inhibition of SARS-CoV-2 replication in the host cell.
In \cite{lin2017effective}, Resveratrol has been demonstrated to be a significant inhibitor MERS-CoV infection. Resveratrol treatment decreases the expression of nucleocapsid (N) protein of MERS-CoV, which is essential for viral replication. 
As MG132 and Resveratrol play a vital role in inhibiting the replication of other coronaviruses SARS-CoV and MERS-CoV, so they may be potential candidates for the prevention and treatment of SARS-CoV-2.  
Another drug Captopril is known as Angiotensin II receptor blockers (ARB), which directly inhibits the production of angiotensin II. In \cite{shang2020structural}, Angiotensin-converting enzyme 2 (ACE2) is demonstrated as the binding site for SARS-CoV-2. So Angiotensin II receptor blockers (ARB) may be good candidates to use in the tentative treatment for SARS-CoV-2 infections \cite{gurwitz2020angiotensin}. 
 
In summary, our proposed method predicts several drug targets and multiple repurposable drugs that have prominent literature evidence of uses as antiviral drugs, especially for two other coronavirus species SARS-CoV and MERS-CoV. Some drugs are also directly associated with the treatment of SARS-CoV-2 identified by recent literature. However, further clinical trials and several preclinical experiments are required to validate the clinical benefits of these potential drugs and drug targets.

\begin{table*}[t]
    \centering
    \caption{Dataset Description}
    \resizebox{\textwidth}{!}{%
    \begin{tabular}{|p{0.2in}|p{1.4in}|p{2.5in}|p{0.7in}|p{1.1in}|p{1.2in}|}
    \hline
    Index & Dataset Category & Dataset & \#Edges & \multicolumn{2}{c|}{\#Nodes}\\
    \hline
     \multirow{2}{*}{1} & \multirow{2}{*}{Human PPI}  & CCSB \cite{yu2011next} &13944  & \multicolumn{2}{c|}{4303}\\
     & & HPRD \cite{peri2003development} & 39240 & \multicolumn{2}{c|}{9617} \\
     \hline
     \multirow{2}{*}{2} & \multirow{2}{*}{SARS-CoV-2--Host PPI}  & Gordon et al\cite{gordon2020sars} &332  & 27 (\#SARS-CoV-2) & 332 (\#Host  ) \\
     & & Dick et al \cite{SP2dick} & 261 & 6 (\#SARS-CoV-2) &  202 (\#Host) \\
     \hline
     3 & Drug-target interaction & DrugBank (v4.3) \cite{law2014drugbank}, ChEMBL \cite{gaulton2012chembl}, Therapeutic Target Database (TTD) \cite{yang2016therapeutic}, PharmGKB database & 1788407 & 1307 (\# Drug) & 12134 (\# Host-target) \\
     \hline
        \end{tabular}%
        }
    
    \label{tab:my_label}
\end{table*}

\begin{longtable}{|p{0.2in}|p{0.4in}|p{1.2in}|p{0.8in}|p{3.4in}|}
\caption{Table shows 92 predicted repurposable drugs}\\
\hline
  Sl. No. & pubchem id & Drug                   & Clinical phase  & Uses                                                       \\ \hline
\endfirsthead
\multicolumn{5}{c}%
{{\bfseries Table \thetable\ continued from previous page}} \\
\endhead
1  & 15281      & Alexidine              & Preclinical     & phosphatidylglycerophosphatase inhibitor                   \\ \hline
2  & 17150      & Apigenin               & Preclinical     & casein kinase inhibitor|cell proliferation inhibitor       \\ \hline
3  & 18998      & Baclofen               & Launched        & benzodiazepine receptor agonist                            \\ \hline
4  & 33216      & Clopamide              & Launched        & sodium/chloride cotransporter inhibitor                    \\ \hline
5  & 37446      & Digoxin                & Launched        & ATPase inhibitor                                           \\ \hline
6  & 107704     & Digitoxigenin          & Preclinical     & ATPase inhibitor                                           \\ \hline
7  & 117069     & Ouabain                & Launched        & ATPase inhibitor                                           \\ \hline
8  & 59105      & Metformin              & Launched        & insulin sensitizer                                         \\ \hline
9  & 71063      & Piribedil              & Launched        & dopamine receptor agonist                                  \\ \hline
10 & 220964     & Pergolide              & Withdrawn       & dopamine receptor agonist                                  \\ \hline
11 & 76299      & Pyrvinium              & Launched        & androgen receptor antagonist                               \\ \hline
12 & 80626      & Sirolimus              & Launched        & mTOR inhibitor                                             \\ \hline
13 & 86324      & Tanespimycin           & Phase 3         & HSP inhibitor                                              \\ \hline
14 & 180858     & Geldanamycin           & Preclinical     & HSP inhibitor                                              \\ \hline
15 & 245370     & Alvespimycin           & Phase 2         & HSP inhibitor                                              \\ \hline
16 & 93086      & Tretinoin              & Launched        & retinoid receptor agonist|retinoid receptor ligand         \\ \hline
17 & 102361     & Anisomycin             & Preclinical     & DNA synthesis inhibitor                                    \\ \hline
18 & 103692     & Atovaquone             & Launched        & mitochondrial electron transport inhibitor                 \\ \hline
19 & 104090     & Azacitidine            & Launched        & DNA methyltransferase inhibitor                            \\ \hline
20 & 104487     & Benperidol             & Launched        & dopamine receptor antagonist                               \\ \hline
21 & 121586     & Sulpiride              & Launched        & dopamine receptor antagonist                               \\ \hline
22 & 104924     & Bisacodyl              & Launched        & laxative                                                   \\ \hline
23 & 105316     & Camptothecin           & Phase 3         & topoisomerase inhibitor                                    \\ \hline
24 & 108903     & Doxorubicin            & Launched        & topoisomerase inhibitor                                    \\ \hline
25 & 114021     & Irinotecan             & Launched        & topoisomerase inhibitor                                    \\ \hline
26 & 115466     & Mitoxantrone           & Launched        & topoisomerase inhibitor                                    \\ \hline
27 & 107326     & Daunorubicin           & Launched        & RNA synthesis inhibitor|topoisomerase inhibitor            \\ \hline
28 & 108095     & Digoxigenin            & Preclinical     & steroid                                                    \\ \hline
29 & 108477     & Diltiazem              & Launched        & calcium channel blocker                                    \\ \hline
30 & 109280     & Emetine                & Phase 2         & protein synthesis inhibitor                                \\ \hline
31 & 118729     & Puromycin              & Preclinical     & protein synthesis inhibitor                                \\ \hline
32 & 178082     & Chlortetracycline      & Launched        & protein synthesis inhibitor                                \\ \hline
33 & 182249     & Midecamycin            & Launched        & protein synthesis inhibitor                                \\ \hline
34 & 199663     & Clindamycin            & Launched        & protein synthesis inhibitor                                \\ \hline
35 & 109651     & Estropipate            & Launched        & estrogen receptor agonist                                  \\ \hline
36 & 110088     & Etamsylate             & Launched        & haemostatic agent                                          \\ \hline
37 & 111183     & Fursultiamine          & Launched        & vitamin B                                                  \\ \hline
38 & 114689     & Medrysone              & Launched        & glucocorticoid receptor agonist                            \\ \hline
39 & 115901     & Nadolol                & Launched        & adrenergic receptor antagonist                             \\ \hline
40 & 116238     & Naftopidil             & Launched        & adrenergic receptor antagonist                             \\ \hline
41 & 119913     & Sotalol                & Launched        & adrenergic receptor antagonist                             \\ \hline
42 & 215415     & Metoprolol             & Launched        & adrenergic receptor antagonist                             \\ \hline
43 & 520283     & Doxazosin              & Launched        & adrenergic receptor antagonist                             \\ \hline
44 & 116649     & Nocodazole             & Preclinical     & tubulin polymerization inhibitor                           \\ \hline
45 & 220050     & Paclitaxel             & Launched        & tubulin polymerization inhibitor                           \\ \hline
46 & 117480     & Parthenolide           & Phase 1         & NFkB pathway inhibitor                                     \\ \hline
47 & 119578     & Scriptaid              & Preclinical     & HDAC inhibitor                                             \\ \hline
48 & 124043     & Vorinostat             & Launched        & HDAC inhibitor                                             \\ \hline
49 & 120719     & Sulconazole            & Launched        & sterol demethylase inhibitor                               \\ \hline
50 & 123302     & Tracazolate            & Phase 2         & GABA receptor modulator                                    \\ \hline
51 & 125857     & Novobiocin             & Launched        & bacterial DNA gyrase inhibitor                             \\ \hline
52 & 129195     & Azacyclonol            & Preclinical     & histamine receptor antagonist                              \\ \hline
53 & 133584     & Galantamine            & Launched        & acetylcholinesterase inhibitor                             \\ \hline
54 & 139192     & Niclosamide            & Launched        & DNA replication inhibitor|STAT inhibitor                   \\ \hline
55 & 145407     & Verteporfin            & Launched        & photosensitizing agent                                     \\ \hline
56 & 150553     & Alprostadil            & Launched        & prostanoid receptor agonist                                \\ \hline
57 & 159244     & Fulvestrant            & Launched        & estrogen receptor antagonist                               \\ \hline
58 & 176723     & Cantharidin            & Launched        & protein phosphatase inhibitor                              \\ \hline
59 & 179449     & Danazol                & Launched        & estrogen receptor antagonist|progesterone receptor agonist \\ \hline
60 & 186039     & MG-132                 & Preclinical     & proteasome inhibitor                                       \\ \hline
61 & 191423     & Ajmaline               & Launched        & sodium channel blocker                                     \\ \hline
62 & 245766     & Ambroxol               & Launched        & sodium channel blocker                                     \\ \hline
63 & 352301     & Cinchocaine            & Launched        & sodium channel blocker                                     \\ \hline
64 & 193762     & Ampyrone               & Preclinical     & cyclooxygenase inhibitor                                   \\ \hline
65 & 203449     & Dirithromycin          & Launched        & bacterial 50S ribosomal subunit inhibitor                  \\ \hline
66 & 212182 & Levonorgestrel & Launched & estrogen receptor agonist|glucocorticoid receptor antagonist|progesterone   receptor agonist|progesterone receptor antagonist \\ \hline
67 & 214498 & Mesalazine     & Launched & cyclooxygenase inhibitor|lipoxygenase inhibitor|prostanoid receptor   antagonist                                              \\ \hline
68 & 217727     & Naringin               & Preclinical     & cytochrome P450 inhibitor                                  \\ \hline
69 & 223747     & Primaquine             & Launched        & antimalarial agent|DNA inhibitor                           \\ \hline
70 & 255979     & Captopril              & Launched        & angiotensin converting enzyme inhibitor                    \\ \hline
71 & 260887     & Chlorzoxazone          & Launched        & bacterial 30S ribosomal subunit inhibitor                  \\ \hline
72 & 268568     & Demeclocycline         & Launched        & bacterial 30S ribosomal subunit inhibitor                  \\ \hline
73 & 270479     & Dihydroergotamine      & Launched        & serotonin receptor agonist                                 \\ \hline
74 & 279244     & Fenoprofen             & Launched        & prostaglandin inhibitor                                    \\ \hline
75 & 291604     & Lobeline               & Launched        & acetylcholine receptor antagonist                          \\ \hline
76 & 293538     & Luteolin               & Phase 2         & glucosidase inhibitor                                      \\ \hline
77 & 295381     & Meclofenoxate          & Launched        & nootropic agent                                            \\ \hline
78 & 295680 & Menadione      & Launched & mitochondrial DNA polymerase inhibitor|phosphatase inhibitor                                                                  \\ \hline
79 & 321706     & Sulfaphenazole         & Launched        & dihydropteroate synthetase inhibitor                       \\ \hline
80 & 333068     & Zardaverine            & Phase 2         & phosphodiesterase inhibitor                                \\ \hline
81 & 372869     & Metronidazole          & Launched        & DNA inhibitor                                              \\ \hline
82 & 416788     & Resveratrol            & Launched        & cytochrome P450 inhibitor|SIRT activator                   \\ \hline
83 & 423536     & Fisetin                & Preclinical     & Aurora kinase inhibitor                                    \\ \hline
84 & 441851     & Morantel               & Launched        & acetylcholine receptor agonist                             \\ \hline
85 & 451754     & Kinetin                & Launched        & cell division inducer                                      \\ \hline
86 & 461005     & Chrysin                & Phase 1         & breast cancer resistance protein inhibitor                 \\ \hline
87 & 471412     & Vinblastine            & Launched        & microtubule inhibitor|tubulin polymerization inhibitor     \\ \hline
88 & 472841     & Meticrane              & Launched        & diuretic                                                   \\ \hline
89 & 483385     & Ethosuximide           & Launched        & succinimide antiepileptic                                  \\ \hline
90 & 494980     & Tetraethylenepentamine & Phase 2/Phase 3 & superoxide dismutase inhibitor                             \\ \hline
91 & 517908     & Ticlopidine            & Launched        & purinergic receptor antagonist                             \\ \hline
92 & 543633     & Todralazine            & Launched        & antihypertensive agent                                     \\ \hline

\label{drug_tab}
\end{longtable}

\begin{table*}[]
\caption{table describing the Gene Ontology (Biological process) and KEGG pathway for 78 CoV-host proteins }
\resizebox{\textwidth}{!}{%
\begin{tabular}{|p{3.5in}|p{1in}|p{2in}|}
\hline
Term (GO/KEGG)                                                          & p-value  & Genes                                     \\ \hline
Herpes simplex infection                                                & 0.002142 & TRAF2, PPP1CA, CASP8, HLA-A, HCFC1, HLA-G \\ \hline
Viral carcinogenesis                                                    & 0.003507 & TRAF2, YWHAG, CASP8, HLA-A, VDAC3, HLA-G  \\ \hline
Endocytosis                                                             & 0.006957 & AP2A2, HLA-A, HSPA6, SMURF1, RAB10, HLA-G \\ \hline
Epstein-Barr virus infection                                            & 0.023579 & TRAF2, HLA-A, TNFAIP3, HLA-G              \\ \hline
Legionellosis                                                           & 0.030506 & EEF1A1, CASP8, HSPA6                      \\ \hline
Viral myocarditis                                                       & 0.033704 & CASP8, HLA-A, HLA-G                       \\ \hline
antigen processing   and presentation (GO:0019882)                      & 8.04E-05 & HLA-H, HLA-A, MR1, RAB10, HLA-G           \\ \hline
antigen processing and presentation of   peptide antigen via MHC class I (GO:0002474)                            & 2.60E-04 & HLA-H, HLA-A, MR1, HLA-G     \\ \hline
negative regulation of extrinsic   apoptotic signaling pathway via death domain receptors (GO:1902042)           & 3.46E-04 & TRAF2, HMOX1, CASP8, TNFAIP3 \\ \hline
antigen processing and presentation of   exogenous peptide antigen via MHC class I, TAP-independent (GO:0002480) & 6.05E-04 & HLA-H, HLA-A, HLA-G          \\ \hline
negative regulation of I-kappaB   kinase/NF-kappaB signaling (GO:0043124)                                        & 6.14E-04 & TRIM59, CASP8, TLE1, TNFAIP3 \\ \hline
cellular response to heat (GO:0034605)                                  & 0.010384 & HMOX1, HSPA6, MYOF                        \\ \hline
nuclear polyadenylation-dependent tRNA   catabolic process (GO:0071038) & 0.020673 & EXOSC8, EXOSC2                            \\ \hline
nuclear polyadenylation-dependent rRNA   catabolic process (GO:0071035) & 0.024756 & EXOSC8, EXOSC2                            \\ \hline
antigen processing and presentation of   exogenous peptide antigen via MHC class I, TAP-dependent (GO:0002479)   & 0.028411 & HLA-H, HLA-A, HLA-G          \\ \hline
type I interferon signaling pathway (GO:0060337)                        & 0.02925  & HLA-H, HLA-A, HLA-G                       \\ \hline
death-inducing signaling complex assembly (GO:0071550)                  & 0.032873 & TRAF2, CASP8                              \\ \hline
low-density lipoprotein particle   clearance (GO:0034383)               & 0.032873 & HMOX1, SCARB1                             \\ \hline
exonucleolytic trimming to generate   mature 3'-end of 5.8S rRNA from tricistronic rRNA transcript (GO:0000467)  & 0.032873 & EXOSC8, EXOSC2               \\ \hline
U4 snRNA 3'-end processing (GO:0034475)                                 & 0.032873 & EXOSC8, EXOSC2                            \\ \hline
interferon-gamma-mediated signaling   pathway (GO:0060333)              & 0.035392 & HLA-H, HLA-A, HLA-G                       \\ \hline
protein processing (GO:0016485)                                         & 0.036307 & PCSK1, TYSND1, PCSK6                      \\ \hline
nuclear-transcribed mRNA catabolic   process, exonucleolytic, 3'-5'(GO:0034427)                                  & 0.036907 & EXOSC8, EXOSC2               \\ \hline
regulation of sequestering of zinc ion (GO:0061088)                     & 0.036907 & SLC30A6, SLC30A7                          \\ \hline
regulation of immune response (GO:0050776)                              & 0.038229 & PVR, HLA-H, HLA-A, HLA-G                  \\ \hline
\end{tabular}%
}
\label{tab:GOtable}
\end{table*}


\section*{Discussion}

In this work, we have successfully generated a list of high-confidence candidate drugs that can be repurposed to counteract SARS-CoV-2 infections. The novelties have been to integrate most recently published SARS-CoV-2 protein interaction data on the one hand, and to use most recent, most advanced AI (deep learning) based high-performance prediction machinery on the other hand, as the two major points. In experiments, we have validated that our prediction pipeline operates at utmost accuracy, confirming the quality of the predictions we have raised.

The recent publication (April 30, 2020) of two novel SARS-CoV-2-human protein interaction resources\cite{gordon2020sars,SP2dick} has unlocked enormous possibilities in studying virulence and pathogenicity of SARS-CoV-2, and the driving mechanisms behind it. Only now, various experimental and computational approaches in the design of drugs against COVID-19 have become conceivable, and only now such approaches can be exploited truly systematically, at both sufficiently high throughput and accuracy.

Here, to the best of our knowledge, we have done this for the first time. We have integrated the new SARS-CoV-2 protein interaction data with well established, long-term curated human protein and drug interaction data. These data capture hundreds of thousands approved interfaces between encompassing sets of molecules, either reflecting drugs or human proteins. As a result, we have obtained a comprehensive drug-human-virus interaction network that reflects the latest state of the art in terms of our knowledge about how SARS-CoV-2 and interacts with human proteins and repurposable drugs. 

For exploiting the new network---already establishing a new resource in its own right---we have opted for most recent and advanced deep learning based technology. A generic reason for this choice is the surge in advances and the resulting boost in operative prediction performance of related methods over the last 3-4 years. A particular reason is to make use of most advanced graph neural network based techniques, namely variational graph autoencoders as a deep generative model of utmost accuracy, the practical implementation of which\cite{salha2020fastgae} was presented only a few months ago (just like the relevant network data). Note that only this recent implementation enables to process networks of sizes in the range of common molecular interaction data. In essence, graph neural networks ``learn'' the structure of links in networks, and infer rules that underlie the interplay of links. Based on the knowledge gained, they enable to predict links and output the corresponding links together with probabilities for them to indeed be missing. 

Simulation experiments, reflecting scenarios where links known to exist in our network were re-established by prediction upon their removal, pointed out that our pipeline does indeed predict missing links at utmost accuracy.

Encouraged by these simulations, we proceeded by performing the core experiments, and predicted links to be missing without prior removal of links in our encompassing network. These core experiments revealed 692 high confidence interactions relating to 92 drugs. In our experiments, we focused on predicting links between drugs and human proteins that in turn are known to interact with SARS-CoV-2 proteins (SARS-CoV-2 associated host proteins). We have decidedly put the focus not on drug - SARS-CoV-2-protein interactions, which would have reflected more direct therapy strategies against the virus. Instead, we have focused on predicting drugs that serve the purposes of host-directed therapy (HDT) options, because HDT strategies have proven to be more sustainable with respect to mutations by which the virus escapes a response to the therapy applied. Note that HDT strategies particularly cater to drug repurposing attempts, because repurposed drugs have already proven to lack severe side effects, because they are either already in use, or have successfully passed the preclinical trial stages.

We further systematically categorized the 92 repurposable drugs into 70 categories based on their domains of application and molecular mechanism. According to this, we identified and highlighted several drugs that target host proteins that the virus needs to enter (and subsequently hijack) human cells. One such example is Captopril, which directly inhibits the production of Angiotensin-Converting Enzyme-2 (ACE-2), in turn already known to be a crucial host factor for SARS-CoV-2. Further, we identified Primaquine, as an antimalaria drug used to prevent the Malaria and also Pneumocystis pneumonia (PCP) relapses, because it interacts with the TIM complex TIMM29 and ALG11. Moreover, we have highlighted drugs that act as DNA replication inhibitor (Niclosamide, Anisomycin), glucocorticoid receptor agonists (Medrysone), ATPase inhibitors (Digitoxigenin, Digoxin), topoisomerase inhibitors (Camptothecin, Irinotecan), and proteosomal inhibitors (MG-132). Note that some drugs are known to have rather severe side effects from their original use (Doxorubicin, Vinblastine), but the disrupting effects of their short-term usage in severe COVID-19 infections may mean sufficient compensation.

In summary, we have compiled a list of drugs, which when repurposed are of great potential in the fight against the COVID-19 pandemic, where therapy options are urgently needed. Our list of predicted drugs suggests both options that had been identified and thoroughly discussed before and new opportunities that had not been pointed out earlier. The latter class of drugs may offer valuable chances for pursuing new therapy strategies against COVID-19.



\section*{Materials and Methods}

\subsection*{Dataset Preparation}
We have utilized three categories of interaction datasets: human protein-protein interactome data, SARS-CoV-2-host protein interaction data, and drug-host interaction data.    

\subsubsection*{SARS-CoV-2-host Interaction Data}
We have taken SARS-CoV-2--host interaction information from two recent studies by Gordon et al and Dick et al  \cite{gordon2020sars,SP2dick}. In \cite{gordon2020sars}, 332 high confidence interactions between SARS-CoV-2 and human proteins are predicted using using affinity-purification mass spectrometry (AP-MS).  In \cite{SP2dick}, 261 high confidence interactions are identified using sequence-based PPI predictors (PIPE4 \& SPRINT).

\subsubsection*{Drug-Host Interactome Data}
The drug–target interaction information has been collected from five databases, viz., DrugBank database (v4.3) \cite{law2014drugbank}, ChEMBL \cite{gaulton2012chembl} database, Therapeutic Target Database (TTD) \cite{yang2016therapeutic}, PharmGKB database, and IUPHAR/BPS Guide to PHARMACOLOGY \cite{pawson2014iuphar}.  Total number of drugs and drug-host interactions used in this study are $1309$ and $1788407$, respectively.

\subsubsection*{The Human Protein–Protein Interactome}
We have built a comprehensive list of human PPIs from two datasets: (1) CCSB human Interactome database consisting of 7,000 genes, and 13944 high-quality binary interactions \cite{rual2005towards,rolland2014proteome,luck2020reference}, (2) The Human Protein Reference Database \cite{peri2003development} which consists of 8920 proteins and 53184 PPIs.\\

The summary of all the datasets is provided in Table \ref{tab:my_label}. CMAP database \cite{subramanian2017next} is used to annotate the drugs with their usage different disease areas.

\subsection*{Sampling Strategy and Feature Matrix Generation}
We have utilized \textit{Node2vec} \cite{grover2016node2vec}, an algorithmic framework for learning continuous feature representations for nodes in networks. It maps the nodes to a low-dimensional feature space that maximizes the likelihood of preserving network neighborhoods.

The principle of feature learning framework in a graph can be described as follows:
Let $G=(V, E)$ be a given graph, where $V$ represents a set of nodes, and $E$ represents the set of edges. The feature representation of nodes ($|V|$) is given by a mapping function: $f: V \rightarrow R^d$, where $d$ specify the feature dimension. The $f$ may also be represented as a node feature matrix of dimension of $|V|\times d$. For each node, $v \in V$, $NN_S(v) \subset V$ defines a network neighborhood of node $v$ which is generated using a neighbourhood sampling strategy $S$. The sampling strategy can be described as an interpolation between breadth-first search and depth-first search technique \cite{grover2016node2vec}. The objective function can be described as:
\begin{equation}
     \max_f \left(\sum_{v \in V} \log P(NN_S(v)|f(v))\right),
\end{equation}
This maximizes the likelihood of observing a network neighborhood $NN_S(v)$ for a node $v$ given on its feature representation $f$. 
Now the probability of observing a neighborhood node $n_i \in NN_S(v)$
given the feature representation of the source node $v$ is given as :
\begin{equation}
    P(NN_S(v)|f(v))= \prod_{n_i \in NN_S(v)}P(n_i|f(v)).
\end{equation}
where, $n_i$ is the $i^{th}$ neighbor of node $v$ in neighborhood set $NN_S(v)$.
The conditional likelihood of each source ($v$) and neighborhood node ($n_i \in NN_S(V)$) pair is represented as softmax
of dot product of their features $f(v)$ and $f(n_i)$ as follows:
\begin{equation}
    P(n_i|f(v))= \frac{exp(f(v).f(n_i))}{\sum_{u \in V}exp(f(u).f(v))}
\end{equation}


\subsection*{Variational Graph Auto Encoder}\label{vgae}
Variational Graph Autoencoder (VGAE) is a framework for unsupervised learning on graph-structured data \cite{rezende2014stochastic}. This
model uses latent variables and is effective in learning interpretable latent representations for undirected graphs. The graph autoencoder consists of two stacked models: 1) Encoder and 2) Decoder. First, an encoder based on graph convolution networks (GCN) \cite{kipf2016variational} maps the nodes into a low-dimensional embedding space.
Subsequently, a decoder attempts to reconstruct the original graph structure from the encoder representations. Both models are jointly trained to optimize the quality of the reconstruction from the embedding space, in an unsupervised way. The functions of these two model can be described as follows:
 
\paragraph{\textbf{Encoder:}} It uses Graph Convolution Network (GCN) on adjacency matrix $A$ and the feature representation matrix $F$. Encoder generates a $d'$-dimensional latent variable $z_i$ for each node $i \in V$, with $|V|=n$, that corresponds to each embedding node, with $d'\le n$. The inference model of the encoder is given below:
\begin{equation}
    r(Z|A,F)= \prod_{i=1}^{|v|} r(z_i|A,F),
\end{equation}
where, $r(z_i|A,F)$ corresponds to normal distribution, $\mathcal{N}(\frac{z_i}{\mu_i},\sigma_i^2)$, $\mu_i$ and $\sigma_i$ are the Gaussian mean and variance parameters. The actual embedding vectors
$z_i$ are samples drawn from these distributions.

\paragraph{\textbf{Decoder:}} It is a generative model that decodes the latent variables $z_i$ to reconstruct the matrix $A$ using inner products with sigmoid activation from  embedding vector, ($Z$). 
\begin{equation}
    \widehat{A}_{i,j}=p(A_{i,j}=1|z_i,z_j)=\,Sigmoid(z_i^T.z_j),
\end{equation}
where, $\widehat{A}$ is the decoded adjacency matrix.
The objective function of the variational graph autoencoder (VGAE) can be written as:
\begin{equation}
    C_{VGAE}= E_{r(Z|A,F)}[\log p(A|Z)]-D_{KL}(r(Z|A,F)||p(Z))
\end{equation}
The objective function $C_{VGAE}$ maximizes the likelihood of decoding the adjacency matrix w.r.t graph autoencoder weights using stochastic gradient decent. Here, $D_{KL}(.||.)$ represents Kullback-Leibler divergence \cite{kullback1951information} and $p(Z)$ is the prior distribution of latent variable.

\subsection*{Drug--SARS-CoV-2 Link Prediction}

\begin{enumerate}
\item \textbf{Adjacency Matrix Preparation} 
In this work, we consider an undirected graph $G = (V, E)$ with $|V|=n$ nodes and $|E|=m$ edges. We denote $A$ as the binary adjacency matrix of $G$. Here $V$ consists of SARS-Cov-2 proteins, CoV-host proteins, drug-target proteins and drugs. The matrix ($A$) contains a total of $n=16444$ nodes given as:
\begin{equation}
    n=|N_{Nc}|+|N_{DT}|+|N_{NT}|+|N_D|,
\end{equation}
where, $N_{Nc}$ is the number of SARS-CoV-2 proteins. $N_{DT}$ is the number of drug targets, whereas $N_{NT}$ and $N_{D}$ represent the number of CoV-host and drugs nodes, respectively.
Total number of edges is given by:
\begin{equation}
    m=|E_1|+|E_2|+|E_3|,
\end{equation}
where, $E_1$ represents interactions between SARS-CoV-2 and human host proteins, $E_2$ is the number of interactions among human proteins, and $E_3$ represents the number of interactions between drugs and human host proteins.
\item \textbf{Feature Matrix Preparation:} The neighborhood sampling strategy is used here to prepare a feature representation of all nodes. A flexible biased random walk procedure is employed to explore the neighborhood of each node.
A random walk in a graph $G$ can be described as the probability:
\begin{equation}
    P(a_i=x|a_{i-1}=v)= \pi(v,x),
\end{equation}
where, $\pi(v,x)$ is the transition probability between nodes $v$ and $x$, where $(v,x) \in E$ and $a_i$ is the $i^{th}$ node in the walk of length $l$.
The transition probability is given by $\pi(v,x)=c_{pq}(t,x)*w_{vx}$, where $t$ is the previous node of $v$ n the walk, $w_{vx}$ is the static edge weights and p, q are the two parameters which guides the walk. The coefficient $c_{pq}(t,x)$ is given by
\begin{equation}
    c_{pq}(t,x) = 
    \begin{cases}
      1/p & \text{distance(t,x) $=0$}\\
      1 & \text{distance(t,x) $=1$}\\
      1/q & \text{distance(t,x) $=2$}
    \end{cases}       
\end{equation}
where, $distance(t,x)$ represents the shortest path distance between nodes $t$ and node $x$.
The process of feature matrix $F_{n \times d}$ generation is governed by the Node2vec algorithm. It starts from every nodes and simulates $r$ random walks of fixed length $l$. In every step of walk transition probability $\pi(v,x)$ govern the sampling. The generated walk of each iteration is included to a walk-list. Finally, the stochastic gradient descent is applied to optimize the list of walks and result is returned. 

\item \textbf{Link Prediction:}
Scalable and Fast variational graph autoencoder (FastVGAE) \cite{salha2020fastgae} is utilized in our proposed work to reduce the computational time of VGAE in large network.
The adjacency matrix $A$ and the feature matrix $F$ are given into the encoder of FastVGAE. The encoder uses graph convolution neural network (GCN) on the entire graph to create the latent representation $(Z)$. 
\begin{equation}
    Z= GCN(A,F)
\end{equation}
The encoder works on full Adjacency Matrix $A$. After encoding, sampling is done and decoder works on the sampled sub graph. 

The mechanism of decoder of FastVGAE is slightly different from traditional VGAE. It regenerate the adjacency matrix $\widehat{A}$ based on a subsample of graph nodes, $V_s$. It uses a graph node sampling technique to randomly sample the reconstructed nodes at each iteration. 
Each node is assigned with a probability $p_i$ and the selection of noes is based on the high score of $p_i$.
The probability $p_i$ is given by the following equation:
\begin{equation}
    p(i)= \frac{f(i)^\alpha}{\sum_{j \in V}f(j)^\alpha},
\end{equation}
where, $f(i)$ is the degree of node $i$, and $\alpha$ is the sharpening parameter. We take $\alpha=2$ in our study. The node selection process is repeated until $|V_s|= n_s$, where $n_s$ is the number of sampling nodes. 

The decoder reconstructs the smaller matrix, $\widehat{A}_s$ of dimension $n_s \times n_s$ instead of decoding the main adjacency matrix $A$. The decoder function follows the following equation:
\begin{equation}
    \widehat{A}_s(i,j)= \,Sigmoid(z_i^T.z_j), \text{  } \forall (i,j) \in V_s \times V_s.
\end{equation}
At each training iteration different subgraph ($G_s$) is drawn using the sampling method.

After the model is trained the drug--CoV-host links are predicted using the following equation:
\begin{equation}
    p(A_{ij}=1|z_i,z_j)=\,Sigmoid(z_i^T,z_j),
\end{equation}
where $A_{ij}$ represents the possible links between all combination of SARS-CoV-2 nodes and drug nodes. For each combination of nodes the model gives probability based on the logistic sigmoid function.


\end{enumerate}



\bibliography{main}



\section*{Acknowledgements}
S.Ray acknowledges support from ERCIM Alain Bensoussan Fellowship programme grant. 
S.Bandyopadhyay acknowledges support from J.C. Bose Fellowship [SB/S1/JCB-033/2016 to S.B.] by the DST, Govt. of India; SyMeC Project grant [BT/Med-II/NIBMG/SyMeC/2014/Vol. II] was given to the Indian Statistical Institute by the Department of Biotechnology (DBT), Govt. of India.
A. Mukhopadhyay acknowledges the support received from the research project grant (Memo No: 355(Sanc.)/ST/P/S\&T/6G-10/2018 dt. 08/03/2019) of Dept. of Science \& Technology and Biotechnology, Govt. of West Bengal, India.


\section*{Availability: All codes and datasets are given in the github link:\\  https://github.com/sumantaray/Covid19}

\end{document}